# Shannon information and integrated information: message and meaning

**Alireza Zaeemzadeh and Giulio Tononi**

Department of Psychiatry, University of Wisconsin, Madison, WI 53719, USA

6001 Research Park Blvd, Madison, WI 53719 USA

Correspondence: gtononi@wisc.edu

**Acknowledgements**. This work was supported by the Templeton World Charity Foundation (TWCF0216). We thank especially Matteo Grasso and Jeremiah Hendren for help with figures and text, and present and past members of the laboratory for helpful discussions.

**Abstract**—Information theory, introduced by Shannon, has been extremely successful and influential as a mathematical theory of communication. Shannon's notion of information does not consider the meaning of the messages being communicated but only their probability. Even so, computational approaches regularly appeal to "information processing" to study how meaning is encoded and decoded in natural and artificial systems. Here, we contrast Shannon information theory with integrated information theory (IIT), which was developed to account for the presence and properties of consciousness. IIT considers meaning as integrated information and characterizes it as a structure, rather than as a message or code. In principle, IIT's axioms and postulates allow one to "unfold" a cause–effect structure from a substrate in a state—a structure that fully defines the intrinsic meaning of an experience and its contents. It follows that, for the communication of meaning, the cause–effect structures of sender and receiver must be similar.

# Introduction

Consider a simple scenario in which Alice sees the house on fire and communicates what she sees and thinks to Bob (Fig. 1). We can assume that, by seeing the house on fire and worrying about her cat, Alice's brain enters a specific state associated with her experience (some neurons firing and some not), after which she texts Bob. When Bob receives the message, it triggers a specific state in his brain, associated with the experience of imagining his house on fire. Clearly, Alice was able to communicate some information about what she saw to Bob, and Bob understood the overall meaning of that information. But how does information convey meaning, and where is the meaning to be found?

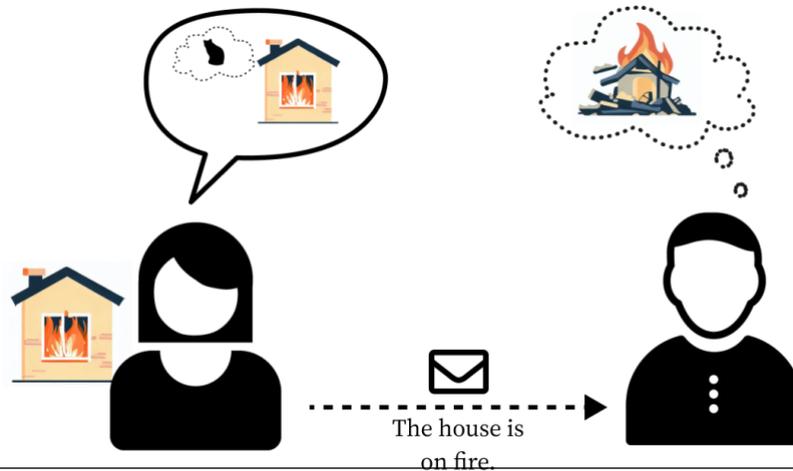

*Fig. 1. Communicating meaning*. *Alice is trying to communicate meaning (conscious contents, such as seeing that the house is on fire) to Bob through a message constituted of symbols. Shannon information theory takes the extrinsic perspective and studies the optimal assignment of symbols and the optimal design of encoding/decoding algorithms. IIT takes the intrinsic perspective of Alice and Bob and studies the intrinsic meaning of system states as conscious contents as well as how meaning can be communicated.*

The term information can be a source of confusion[1] and its relationship to meaning, another loaded term[2, 3], can be problematic. Here, we focus on two fundamental uses of the term: information as *messages* encoding symbols for transmission, processing, and storage, studied by Shannon information theory[4], and information as *meaning*, understood as a structure corresponding to a conscious content, addressed by integrated information theory (IIT)[5]. Both Shannon information theory and IIT start from axioms and employ the tools of probability theory, but their formalism and definition of information are different and complementary.

Information theory quantifies the information content of the message as reduction of uncertainty from the extrinsic perspective of a communication engineer. It provides the formalism for encoding, transmitting, processing, and decoding the message optimally from source to target over a (noisy) channel. As recognized by Shannon himself, it has nothing to say about the meaning of the messages being communicated. IIT, in contrast, quantifies the information content of the experience triggered by the message from the intrinsic perspective of a conscious subject. It provides a formalism to characterize the intrinsic meaning of the experience as a cause–effect structure unfolded from the substrate of the subject's consciousness. In short, Shannon information theory deals with messages and their extrinsic transmission as codes, whereas integrated information deals with meanings and their intrinsic existence as structures. This distinction is important when it comes to the brain, especially in view of the pervasive reference to "information processing" and "codes" in neuroscience and psychology.

In what follows, we briefly introduce some relevant notions from both information theory and IIT, explained in more detail in the *Supplementary material*. Next, we consider side-by-side Shannon information and integrated information and highlight some of their differences. We then return to Alice and Bob and ask how the meaning of what she sees (and what he imagines) is instantiated in their brain's activity patterns. Finally, we consider how meaning can be communicated and the respective roles of Shannon information and integrated information.

## Shannon information theory

The mathematical theory of communication introduced by Claude Shannon, also known as *information theory*, is concerned with reliably communicating symbols over noisy channels through optimal coding and computational overhead (Fig. 2)[4]. Shannon started with a set of desired properties for an information measure, which he called entropy. In a standard formulation, the "axioms" of Shannon information are continuity with respect to the probabilities, additivity, monotonic increase with number of outcomes, maximality for equiprobable outcomes, and symmetry under the permutation of symbols. Information can then be quantified as the number of binary digits (bits) generated by a source (the *source entropy*) and/or successfully transmitted over a channel (the *mutual information*). A channel's *capacity* is the maximum rate at which information can be transmitted through it (the maximum of mutual information between source and target). Measures of information satisfying these properties are employed in virtually every area of modern science and society, and are essential to communication engineering, signal processing, machine learning, and neuroscience[6, 7].

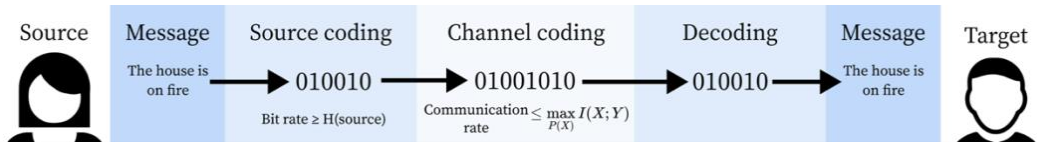

*Fig. 2. Information theory and the communication of symbols. In information theory, the information content of a message is quantified as the minimum number of uniquely decodable symbols (binary digits) assigned to it. In a communication system, unwanted redundancies are removed (source coding) and errors are corrected by adding communication and computation overhead (channel coding). The source-target pair can achieve optimal communication by utilizing the optimal encoding/decoding algorithms designed by an external observer.*

Messages (sequences of symbols) transmitted across a channel are encoded and decoded: a *code* is simply a mapping from one set of symbols to another. Shannon proved that it is always possible to encode and decode messages such that transmission is lossless and as efficient as possible. In general, some information will be transmitted as long as there is some statistical dependence between source and target, implying that some question about the source can be answered based on a readout of the target. Importantly, one can design channels such that they extract some "relevant variable" that conveys information about specific questions and ignores others. This can be considered as a form of *information processing*[8]. However, information processing can only ever decrease the information transmitted over a channel[8].

For Shannon, the "information content" of a symbol or code is purely a function of its probability of occurrence: regardless of how the symbol is interpreted, the less frequent it is, the higher its information content. As he readily recognized[4], information theory is not concerned with the meaning of a symbol: that must be provided by the source and the target or by an external observer who interprets the message or its origin and consequences. *Supplementary material I* offers a brief summary of some relevant aspects of information theory. For an in-depth treatment, see, for example, [8].

## Integrated information theory

*Integrated information theory* (IIT) is concerned with determining the quality and quantity of consciousness from the causal properties of a substrate[5] (Fig. 3). To be conscious is to have an experience—seeing a face, hearing a voice, having a thought, witnessing a house on fire, imagining it, or dreaming it. In other words, you are conscious if "there is something it is like to be you"[9]. IIT aims to account for the properties of experience—which is subjective—in objective, physical terms.[10]

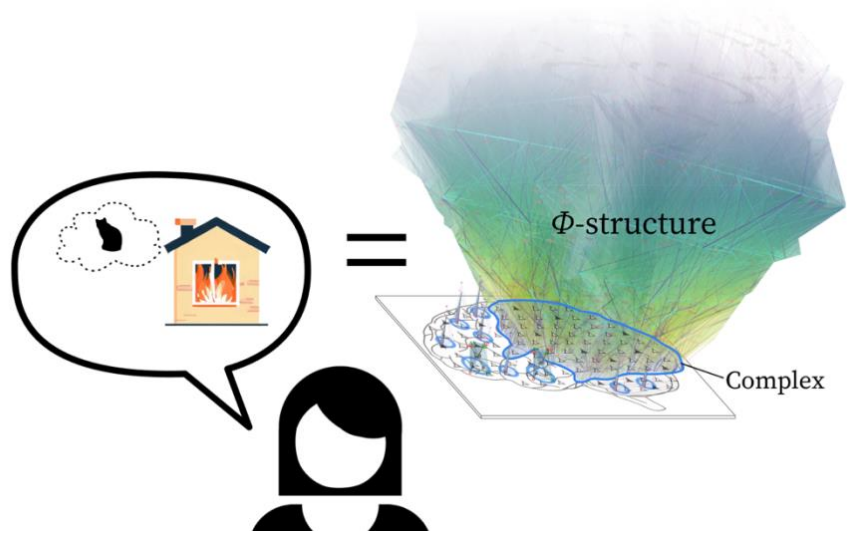

*Fig. 3. IIT and the structure of meaning. IIT defines information as meaning, understood as the content of experience (the feeling is the meaning). It takes the intrinsic perspective of the system and characterizes the meaning of an activity pattern as the Φ-structure unfolded from a complex (a maximally irreducible substrate). The main complex is outlined in blue on the surface of the cerebral cortex (assuming the maximum of irreducibility is over posterior areas), at the grain of individual neurons. The Φ-structure unfolded from it is depicted by a set of vertices (distinctions) bound by edges (relations). Other regions of the brain, such as the cerebellum and parts of the cerebral cortex, can only support very small complexes and associated Φ-structures because they are organized in a modular manner.*

The starting point of IIT is the existence of experience—its "zeroth" axiom. IIT then characterizes the essential properties of experience—those that are immediately and irrefutably true of every conceivable experience—as the five axioms of phenomenal existence[3]:

(1) Intrinsicality. Every experience (say, Alice's experience of seeing the house on fire), is *subjective—for the experiencer*, from her intrinsic or private perspective. It cannot be "for no one."

(2) Information. It is *specific*—this one, the house on fire. It cannot be "generic."

(3) Integration. It is *unitary*—a whole (the house on fire), irreducible to separate experiences. It cannot be "multiple" (one of a house and another of a fire).

(4) Exclusion. It is *definite*—this whole, containing all it contains, neither less nor more. It cannot be "indefinite."

(5) Composition. It is *structured*—composed of distinctions and the relations that bind them together and make it feel the way it feels (the house and its contours, the fire and its flames, all bound to a spatial location, to various colors, and so on). It cannot feel "in no way." The way the experience feels is also what it means, which is therefore fully intrinsic. In short, *the feeling is the meaning*[11].

The next step of IIT is to formulate the essential phenomenal properties of experience (the axioms) in terms of corresponding physical properties of its substrate (the postulates). Physical existence is defined as *cause–effect power* of a substrate of units, understood in operational terms of manipulation and observation: "if we do this" (manipulation—set an initial state), "we see that" (observation—read out the ensuing state). A substrate can thus be fully characterized by its transition probability matrix. On this basis, IIT provides a mathematical formalism to identify a substrate of consciousness and the quality and quantity of experience it supports. The substrate must have cause–effect power whose properties mirror the axioms above: it must be *intrinsic* (for itself), *specific* (in its current, specific state, with a specific cause and effect state), *unitary* (irreducible to its parts), and *definite* (the set of units that is maximally irreducible). Such a substrate is called a *complex* and its irreducibility is measured by system integrated information ($\varphi_s$). Finally, the cause–effect power of the complex must be *structured*, by causes and effects specified by subsets of its units (distinctions) and by their overlaps (relations). The *cause–effect structure* (or $\Phi$-*structure*) unfolded from a complex in its current state fully characterizes, with no additional ingredients, how an experience feels—its quality—which defines its intrinsic meaning. The quantity of feeling/meaning is measured by $\Phi$—the sum of integrated information of the distinctions and relations composing the $\Phi$-structure. A content of experience is then a sub-structure within a $\Phi$-structure.

IIT has strong explanatory and predictive power. It can account for why certain brain regions support consciousness and others do not[12], why consciousness fades during dreamless sleep[12], and why different kinds of experiences feel the way they do, such as extended in space and flowing in time[13-15]. It leads to counterintuitive predictions that are currently being tested[16], to practical applications[12, 17, 18], and to inferences concerning the presence or absence of consciousness in natural and artificial systems, such as computers[19, 20]. *Supplementary material II* offers a brief summary of IIT's concepts, which are presented in detail in "IIT 4.0"[5] and in an online wiki[21].

# Shannon information and integrated information side-by-side

The distinctive approach and goals of information theory and IIT are reflected in several differences between Shannon information and integrated information. It is useful to illustrate these differences with simple examples that demonstrate the requirements imposed by IIT's postulates (Table 1 and Fig. 4).

("0$^{th}$") *Correlational vs. causal*. Shannon information requires a correlation between source and target; integrated information requires cause–effect power. In Fig. 4(a), units *A* and *B* convey information in the Shannon sense (about one another, being correlated, and about unit *X*) but do not have any effects of their own. However, by the 0$^{th}$ postulate of IIT, a candidate system must have cause–effect power—the ability to "take and make a difference," as assessed through interventional probabilities.

(1$^{st}$) *Extrinsic vs. intrinsic*. Shannon information can be between any source and any target; integrated information requires intrinsic cause–effect power. Fig. 4(b) shows how Shannon information can be defined over any set of source and target units. However, IIT's intrinsicality postulate requires that the cause–effect power of a candidate system be over itself.

(2$^{nd}$) *Generic vs. specific*. Shannon information is defined over an ensemble of symbols as the entropy function. Similarly, the mutual information between subsets of units measures the statistical dependence between random variables, which is state-independent (Fig. 4(c)). By IIT's information postulate, however, the cause–effect power of the system must be specific: the substrate must be in a specific state, with these units 'on' and those 'off,' and select a specific cause and effect state over itself.

(3$^{rd}$) *Segregated vs. integrated*. Shannon information benefits from segregation; integrated information requires the substrate's irreducibility. Fig. 4(d) shows how, due to the additivity axiom of information theory, mutual information adds up when a segregated source–target pair is added. On the other hand, by IIT's integration postulate, a candidate system must be causally irreducible ($\varphi_s>0$). Therefore, causally segregated units cannot belong together and do not increase the amount of information specified by a substrate over itself. Indeed, the integrated information of a set of causally independent units is zero.

(4$^{th}$) *Additive vs. definite*. Shannon information benefits from a larger channel; integrated information requires a maximally irreducible substrate with a definite border and grain. As shown in Fig. 4(e), in information theory, the channel, source, and target units (and grain) over which symbols should be transmitted and read out can be chosen based on extrinsic convenience. Also, the additivity of Shannon information implies that larger (and finer) sources and channels are always potentially more informative. By contrast, in IIT, the exclusion postulate requires that a complex must have a definite border (and grain), which is defined by maximal irreducibility and is thus non-arbitrary. This border excludes any larger or smaller entity constituted of overlapping units. In fact, larger systems may have lower $\varphi_s$ values than some of their subsets.

(5$^{th}$) *Holistic vs. structured*. Finally, Shannon information is not structured; integrated information is. Fig. 4(f) shows that, in information theory, subsets of the system do not transmit information beyond what is transmitted by the whole. Statistical dependencies among units are considered as redundancy and do not contribute to the information content of the source. In fact, such redundancies are exploited in source coding to obtain a more compressed representation of the symbols. By contrast, IIT's composition postulate implies that the subsets of a complex contribute information content by structuring its cause–effect power as distinctions (cause–effects of subsets) and relations (overlaps among causes and/or effects), which together compose the $\Phi$-structure of the complex.

Ultimately, because Shannon information is not structured intrinsically, the meaning of symbols (codes) transmitted over a channel must be provided extrinsically, as explicitly recognized by Shannon himself.[1] In IIT, instead, the meaning of an activity pattern is given by the *Φ*-structure unfolded from the complex in that state, that is, intrinsically.

| IIT postulate | Information theory | Integrated information theory |
|---|---|---|
| Existence (0<sup>th</sup>) | *correlational vs. causal* | |
| | Evaluates statistical dependencies through observations. | Evaluates causal interactions through interventions. |
| Intrinsicality | *extrinsic vs. intrinsic* | |
| | Evaluates statistical dependencies between a source and a target. | Evaluates cause–effect power within a candidate system. |
| Information | *generic vs. specific* | |
| | Evaluates statistical dependencies between distributions of symbols, i.e., random variables. | Evaluates the cause–effect power of a specific system state over a specific system state. |
| Integration | *segregated vs. integrated* | |
| | Information can be provided by independent units. | Information must be irreducible to that provided by independent units. |
| Exclusion | *additive vs. definite* | |
| | Adding more units never decreases the information. | Integrated information is maximal for a definite set of units. |
| Composition | *holistic vs. structured* | |
| | Information is not structured, corresponding to the optimal number of binary digits generated or transmitted. | Information is structured, composed of causal distinctions bound by causal relations. |
| **Meaning** | Must be provided extrinsically by an observer who can interpret a code. | Is defined intrinsically by the *Φ*-structure unfolded from a complex in a state. |

***Table I. Shannon information vs. integrated information following IIT's postulates.***

---

[1] Likewise, similar notions of information or generalizations, such as Kolmogorov mutual information and partial information decomposition, are not aimed at unfolding the intrinsic structure of a system's state.

Kolmogorov complexity is the algorithmic complexity of a symbol, defined as the length of the shortest program (in bits) that generates that symbol and halts.[22] Furthermore, conditional Kolmogorov complexity can be defined as the length of the shortest program that takes a symbol as input and generates another symbol and halts. Kolmogorov mutual information is the difference between Kolmogorov complexity and conditional Kolmogorov complexity. Similar to Shannon's measures, such a notion of information can only assign meaning to symbols from the perspective of an external observer and through an extrinsic variable, that is, the algorithmic implementation.

Partial information decomposition[23] generalizes the tools of information theory by defining partial information variables. The mutual information that a set of source variables provides about a given target variable is decomposed into unique information, redundancy, and synergy. For the special case of two source variables, the unique information corresponds to the information that one variable provides and the other does not, the redundant information is the information both of the source variables provide, and the synergistic information is the information that the combination of both variables provides, which is not available from each of the variables alone. The notion of partial informationvariables is consistent with Shannon information (redundancy can coincide with mutual information in special cases) and is an extrinsic quantity defined and measured by an observer outside the system. More importantly, by itself, information decomposition yields a non-negative decomposition into partial information atoms and an ordering among them[23]; it does not yield a structure—that is, a set of objects and the relations among any subset of them.

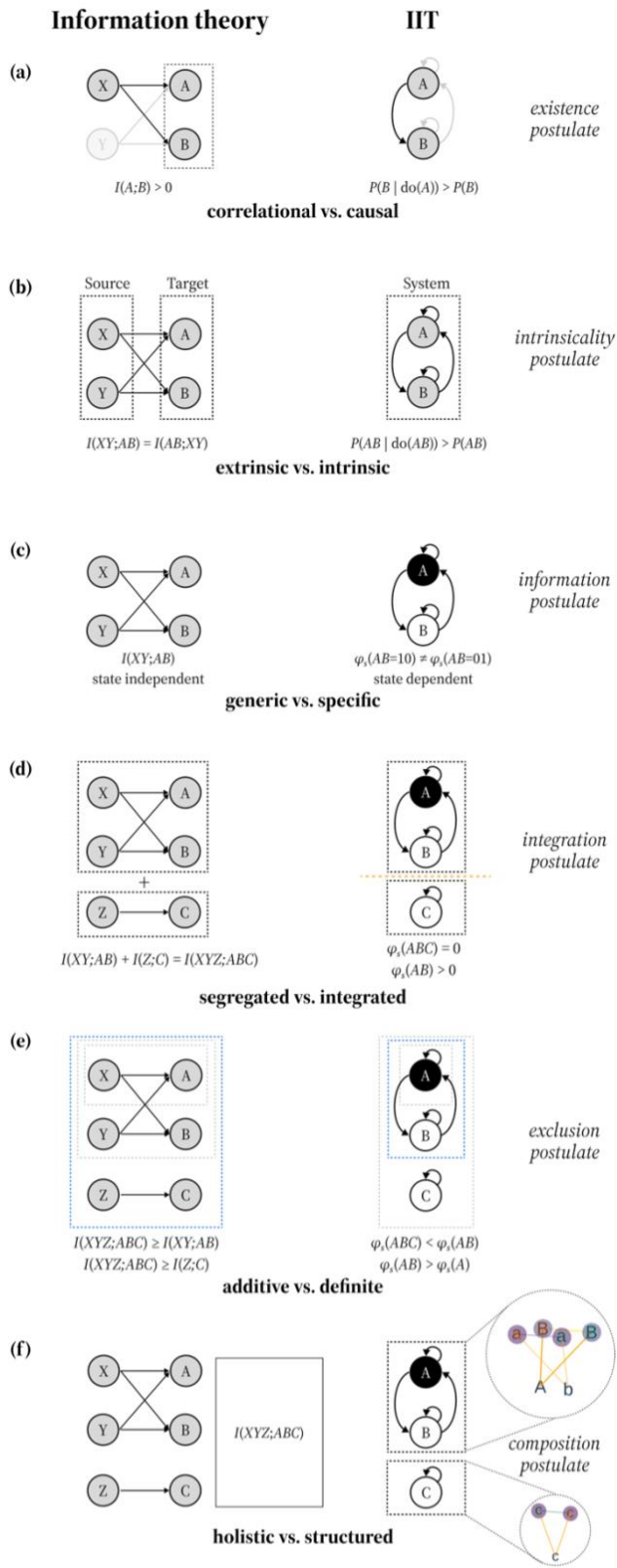

*Fig. 4. Differences between Shannon information and integrated information.* (a) Shannon information captures correlations (e.g., between units A and B owing to their common input from unit X). In contrast, integrated information captures cause-effect power—the ability of units to "take and make a difference," as assessed through interventional probabilities (e.g., using the "do-operator" between units A and B). (b) Shannon information concerns the extrinsic correlations between a source and a target, while integrated information is intrinsic in that it captures causal power of a system over itself. (c) Shannon information is generic (state independent, gray units), while integrated information is always specific (state dependent [black = ON, white = OFF]). (d) Shannon information is additive for causally segregated units, while integrated information is not (as measured across a minimal partition, orange dotted line). (e) Shannon information is non-decreasing with the number of units or channels, while integrated information is maximal over a definite set of units (often not the largest set). (f) Shannon information is holistic (subsets of the system do not specify information beyond the whole), while integrated information is structured (subsets of the system specify causal distinctions and relations not captured by the whole).

## Information processing, coding, and meaning

To appreciate the difference between Shannon information and integrated information with respect to meaning, let us now return to Alice and Bob. Neuroscience and psychology have often borrowed the language of information theory and treated the brain as a biological "information-processing" device that performs sophisticated computations in the service of various functions[7].

Neuroscientists frequently talk of "cracking the neural code"[24]. To exemplify, consider the schematic depiction in Fig. 5 (left panel). Seeing the house on fire triggers a complex set of neuronal interactions in Alice's brain. An activity pattern over Alice's retina is transmitted through thalamic nuclei and triggers an activity pattern in primary visual cortex (V1), which could be considered as a symbol or code—encoding, for example, the location of a stimulus in space. From there, divergent feed-forward channels convey information to a set of higher-level cortical areas (say, V3 and V4). Each area "processes" the information through several "computations," extracting a different variable that may correspond to a relevant category (see *Supplementary material I*). For instance, activity patterns in one area might encode the variable "house" vs. "no house," a code that can be passed on to other areas. Information converging from V3 and V4 would then be "integrated" by a downstream area that computes a code for "danger" vs. "no danger." In line with the information-processing metaphor, one could also envision the brain as performing the equivalent of error correction. For example, top-down codes from high-level areas carrying "contextual" information could counteract noise and disambiguate the message conveyed by the stimulus.[2] Moreover, stimulus information could be complemented by endogenous sources of information, say, encoding the thought "worried about the cat" and the intention "should text Bob." In some views, a competition among local codes would lead to a winning code, which would be broadcast globally throughout the brain and would correspond to information that is consciously accessible[28, 29]. Pre-motor and motor areas could then decode the winning code and the intention, leading to a motor output. Ideally, one would be able to explain what the brain does in terms a set of computations carrying out information processing on assorted neural codes to perform multiple adaptive functions, such as stimulus location, coordinate transformation, categorization, representation, attentional selection, memorization, and valuation, as well as goal selection, decision making, and so on (Fig. 5, left panel).

The information processing approach to the brain and its functions has been successful in explaining many aspects of how the brain does what it does. However, its shortcomings should not be overlooked. One is that codes, computations, and functions do not map tidily onto brain circuits the way they might onto a computer. As a biological system shaped by evolution, development, and learning, the brain is marked by pleiotropy (multiple effects produced by the same cause) and degeneracy (multiple causes producing the same effect)[30]. It is thus unlikely to comply with neat functional subdivisions. The brain is also unique in its extraordinary connectivity, coupled with spontaneous activity and plasticity. It is not surprising, then, that information about sensory inputs, motor plans, task settings, expectations, and reward can be decoded in a highly distributed manner, from the cerebellum to the frontal pole[31]. Moreover, being able to decode variables or contents that are meaningful to us does not mean that those patterns are meaningful from the

---

[2] Alternatively, the brain might work through "predictive processing," sending messages top-down to sensory areas[25-27]. These "inferences" about stimuli would then be updated based on error signals provided by stimulus information.

intrinsic perspective of the brain[32-34].[3]

Then there is the issue of which substrate one should "decode." The brain does not qualify as a channel with well-defined sources and receivers. Should one decode activity patterns from the brain as a whole, the cerebral cortex, the cerebellum, or smaller sub-regions? Based on what criteria? Should one care whether the readout is from a substrate that is highly integrated, and why, given that information benefits from independence? And at which grain should one decode? Over individual neurons and spikes, distributed populations and mean firing rates, or individual synapses and calcium release?[4] Finally, by what criteria would all these different codes be related to one another?

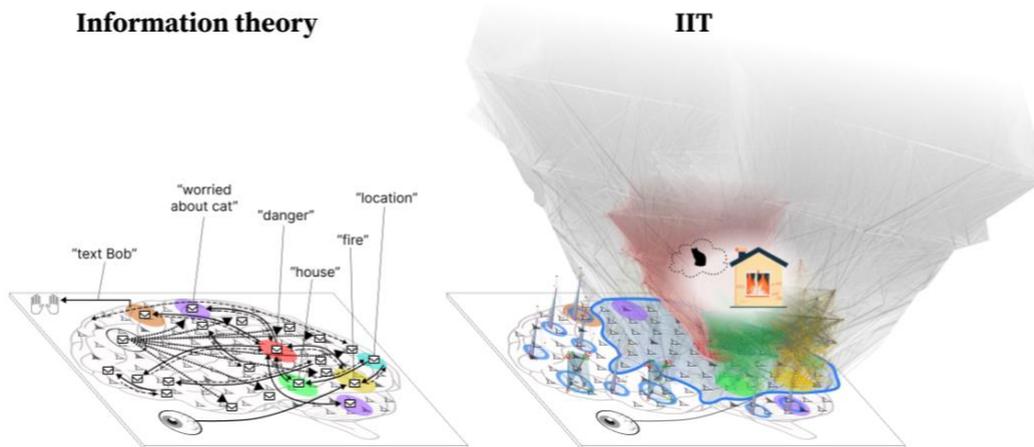

*Fig. 5. Information processing and information structures in the brain. Left panel: the brain as an information-processing device. Information in the Shannon sense is extracted by different circuits, processed, transferred, integrated, and broadcast across many areas. It can be decoded both locally and globally at many different grains. The interpretation of the code is provided extrinsically, by an outside observer, based on predefined categories or on assumptions about the computations/functions performed by various neural circuits. Right panel: the brain as a substrate for information structures. IIT identifies the main complex and its grain (say, the region inside the blue border, at the grain of neurons, over one hundred milliseconds). In its current state, the main complex supports a Φ-structure (gray shape) composed of integrated information: subsets of neurons contribute different contents of experience as Φ-folds (highlighted in different colors) related among themselves. The Φ-structure fully defines the intrinsic feeling/meaning of the experience. Groups of neurons outside the main complex interact with it and among themselves (hence Shannon information can be decoded from them) but constitute very small complexes supporting trivial Φ-structures.*

---

[3] After all, we could decode the time of day from the length or orientation of a tree's shadow, but that does not mean that it encodes time for the tree.

[4] Note that an observer could in principle map any program onto almost any thermodynamically open substrate through an appropriate encoding of inputs and outputs[35, 36].

## Information processing vs. information structures

But the most important shortcoming of information processing approaches has to do with meaning. Even if one were able to sensibly decode activity patterns from different circuits based on plausible computations and functions one might project onto those circuits, where would meaning come from? Ultimately, information processing is a mapping between input and output patterns. As emphasized by Shannon, if an activity pattern is treated as a symbol, it will have no meaning on its own. Its meaning must be attributed *extrinsically*, *relative* to observers that provide their own meaning to what the brain (or a computer) might do. Yet when Alice sees the house on fire and worries about her cat, the feeling/meaning of her experience is *intrinsic* (for her), and *absolute*, rather than relative to an external observer who interprets what she does. And the meaning is present for her "here and now," regardless of what the environment might be like and of what her brain might have done or do next.

Here is where IIT's characterization of feeling/meaning as integrated information can explain how a specific activity pattern can have intrinsic meaning. This is because IIT considers information not as a mapping between input and output patterns, but as a structure—a $\Phi$-structure supported by a complex in a state.

As we have seen, IIT's axioms capture the essential properties of experience, which can be formulated operationally as properties of cause–effect power—IIT's postulates. The first four postulates require that, because experience is intrinsic, specific, unitary, and definite, the substrate of experience should be a maximum of intrinsic, specific, irreducible cause-effect power. In principle, this allows one to determine, from a system's transition probability matrix and its current state, the border and grain of the main complex (the largest complex over a substrate). For example, as illustrated schematically in Fig. 5 (right panel), this might correspond to a definite set of units (localized primarily over posterior cortical regions), of a definite grain (say, neurons), which are either active or not over a definite interval (say, thirty milliseconds).

Furthermore, because experience is structured by phenomenal distinctions and relations that make it feel the way it feels, the composition postulate requires that the cause-effect power of the main complex be unfolded into causal distinctions and relations that compose its $\Phi$-structure. For example, within the main complex, subsets of neurons contribute different contents of experience as sub-structures of the $\Phi$-structure ($\Phi$-folds), such as "house," "fire," and "danger" (highlighted in different colors), which are further related among themselves. According to the theory, the $\Phi$-structure defines in full, with no additional ingredients, the feeling/meaning of the experience.[5] The feeling/meaning is defined intrinsically, *by* the complex and *for* the complex, and in *absolute* terms, "here and now," solely through the distinctions and relations specified by its subsets,[6] with no reference to how an outside observer might interpret it, to the performance of some function or computation, to its similarity to other experiences, to what might happen next, or to the environment.[7] In short, information is a structure, not a symbol or code, a process, a computation, or a function.[8]

---

[5] The composition of a $\Phi$-structure, for a specific activity pattern, is determined by the intrinsic connectivity of the main complex. To a large extent, the latter is organized the way it is because it has adapted to a changing environment during a long evolutionary, developmental, and learning history. In this sense, the environment is partially responsible for the meaning of an experience, but only indirectly so—historically, rather than here and now.

[6] Along the same lines, the "integration of information" is not the computation of an output by a brain region merging multiple inputs, but a large $\Phi$-fold—the set of the distinctions and relations jointly specified by multiple regions.

While an activity pattern on Alice's retina eventually triggers an activity pattern in her main complex, which specifies a $\Phi$-structure whose feeling/meaning is intrinsic and absolute, activity patterns elsewhere in Alice's brain would not be associated with any feeling/meaning, or only minimally so. In Fig. 5 (right panel), this is illustrated by a multitude of very small complexes scattered over the rest of the brain, supporting $\Phi$-structures of negligible $\Phi$ but interacting among themselves in complicated ways. For example, circuits in Alice's hypothalamus and brainstem constantly monitor and regulate blood pressure—they are clearly processing blood pressure information. Yet blood pressure regulation goes on completely "in the dark"—it does not have any feeling/meaning for her. Or consider the cerebellum. Like the cortex, the cerebellum is connected to various sensory and motor pathways (and, indirectly, to the cortex itself). It "represents" many aspects of the body and the environment and is involved in many complex processes. In fact, from cerebellar activity patterns, it is possible to decode not only sensory and motor variables, but also cognitive and emotional ones[42]. Yet the cerebellum, despite having four times more neurons than the cerebral cortex, can be completely removed without appreciably changing the feeling/meaning of what the person experiences[43].[9] So, an activity pattern in certain parts of Alice's cerebral cortex signifies "house on fire, worried about the cat" intrinsically, but concurrent activity patterns in her hypothalamus, brainstem, and cerebellum signify nothing.[10]

Equating meaning with feeling also implies that, without the feeling, there would be no meaning at all. Without subjects who can consciously observe and interpret the world and conceive of functions and computations, there would be no meaning whatsoever. Just imagine a world devoid of conscious beings but populated by artificial intelligence (AI), implemented on digital computers, that can perform any task as well or better than we do. According to IIT, digital computers have the wrong architecture to be able to support complexes of high $\Phi$ and therefore consciousness[19]. In such a world everything would happen "in the dark:" nobody would see anything, hear anything, or think anything.[11] A

---

[7] This view of meaning also differs from extrinsic views that assign meaning based on the similarities among activity patterns or their location within "semantic spaces"[37-41].

[8] If one were to insist on the coding metaphor, we could say that IIT provides the ultimate recipe for decoding intrinsic meaning: it specifies over which substrate to decode (the main complex), at what grain (the one maximizing the $\varphi$ value of the complex), and how to unfold (decode) its $\Phi$-structure. It also provides a way to characterize the contribution to intrinsic meaning of an individual unit (say, an activated neuron in a "face area") through the $\Phi$-fold it specifies (the set of distinctions it contributes, alone and in combination with other units, and the associated relations).

[9] This is consistent with IIT's notion that the cerebellum, owing to its strictly modular, feed-forward architecture, cannot support a complex of high $\Phi$, unlike the cerebral cortex, much of which is organized like a dense lattice[12].

[10] In fact, even activity patterns in certain parts of cortex may also signify nothing if the underlying organization of connections is highly modular, as seems to be the case for much of prefrontal cortex. Similarly, patterns of activity in posterior cortex may signify nothing during dreamless deep sleep, even though neurons are typically still active. This is because breakdown of causal interactions due to neuronal bistability leads to the disintegration of the cortical complex[12]. Finally, some neuronal activity in posterior cortex may signify nothing even during wakefulness, if it specifies cause-effect states that are incongruent (different) with respect to the cause-effect state specified by the main complex as a whole. This may occur, for example, during binocular rivalry, when many neurons "encode" information about an unseen stimulus that is incongruent with the seen stimulus.

[11] Of course, if we presuppose that consciousness is nothing but some computation or function, then computers mimicking us would be conscious by definition. If IIT is right, however, computers can be

description of a visual scene, such as a "house on fire," encoded in a high-level language, would be decodable from some portion of the computer memory inside an Alice-like machine. But this high-level code and associated computations would have no intrinsic meaning for the device, because it would have no feeling.[12]

## Communicating integrated information as meaning

IIT's account of the intrinsic meaning of an activity pattern as a $\Phi$-structure composed of integrated information has direct implications for its communication. In short, meaning can only be communicated if a $\Phi$-fold within a $\Phi$-structure specified by a source complex triggers a similar $\Phi$-fold within a $\Phi$-structure specified by a target complex. The requirements for the communication of integrated information as meaning are thus much more stringent than those for the communication of symbols across a channel.

In Fig. 6(a), two computer programs endowed with artificial intelligence send symbols back and forth, performing computations on the symbols that could be interpreted as functionally equivalent to a conversation between two humans. However, per IIT, standard computers have an internal architecture that is unsuitable for supporting $\Phi$-structures of high $\Phi$; hence they would experience nothing[5, 19], and their computations would have no intrinsic meaning.[13] In this scenario, the communication of Shannon information is excellent, but there is no intrinsic meaning to communicate.

The opposite scenario is depicted in Fig. 6(b). Here, both the source complex and the target complex support $\Phi$-structures of high $\Phi$ that are intrinsically meaningful. However, owing to the channel being blocked, symbols sent by the source complex cannot trigger intended $\Phi$-folds at the target complex. Because Shannon information across the channel is zero, no meaning can be communicated.

Fig. 6(c) illustrates a more interesting scenario. In this case, the source and target complexes support $\Phi$-structures of high $\Phi$, and the symbols are communicated perfectly across the channel; hence Shannon information is high. Even so, little meaning is communicated because the symbols sent from the source complex fail to trigger the intended $\Phi$-fold in the target complex. This may happen, for example, because the source and target speak different languages. In this case, the symbols transmitted are perceived at a shallow level (say, as letters or phonemes), but they do not percolate deep into the target complex and thus trigger few distinctions and relations. Therefore, perceptual richness is low (*Supplementary material II* and [11]) and little meaning is communicated. More generally, even if source and target speak the same language, if their internal architecture is substantially different, communication of meaning will be reduced. There may also be outright miscommunication, in the sense that the message may trigger in the target a meaning radically different from the one intended by the source.

---

functionally equivalent to us but cannot be conscious because, like in the case of the cerebellum, their substrate cannot support $\Phi$-structures of high $\Phi$[19]. Crucially, IIT's validity can be assessed empirically by determining whether the theory can account for the presence and quality of our own consciousness and how it relates to its neural substrate.

[12] The Chinese room argument[44], which was originally aimed at showing that syntax (processing) is not enough for semantics (meaning), shares a similar insight. In the argument, one imagines being conscious and performing a set of functions correctly—answering questions in Chinese—but without any awareness of their meaning. In a world with no conscious beings, all kinds of functions could be carried out without anybody experiencing anything, and therefore without any intrinsic meaning.

[13] Strictly speaking, small subsets of transistors in each computer may have trivially low but non-zero $\Phi$[19].

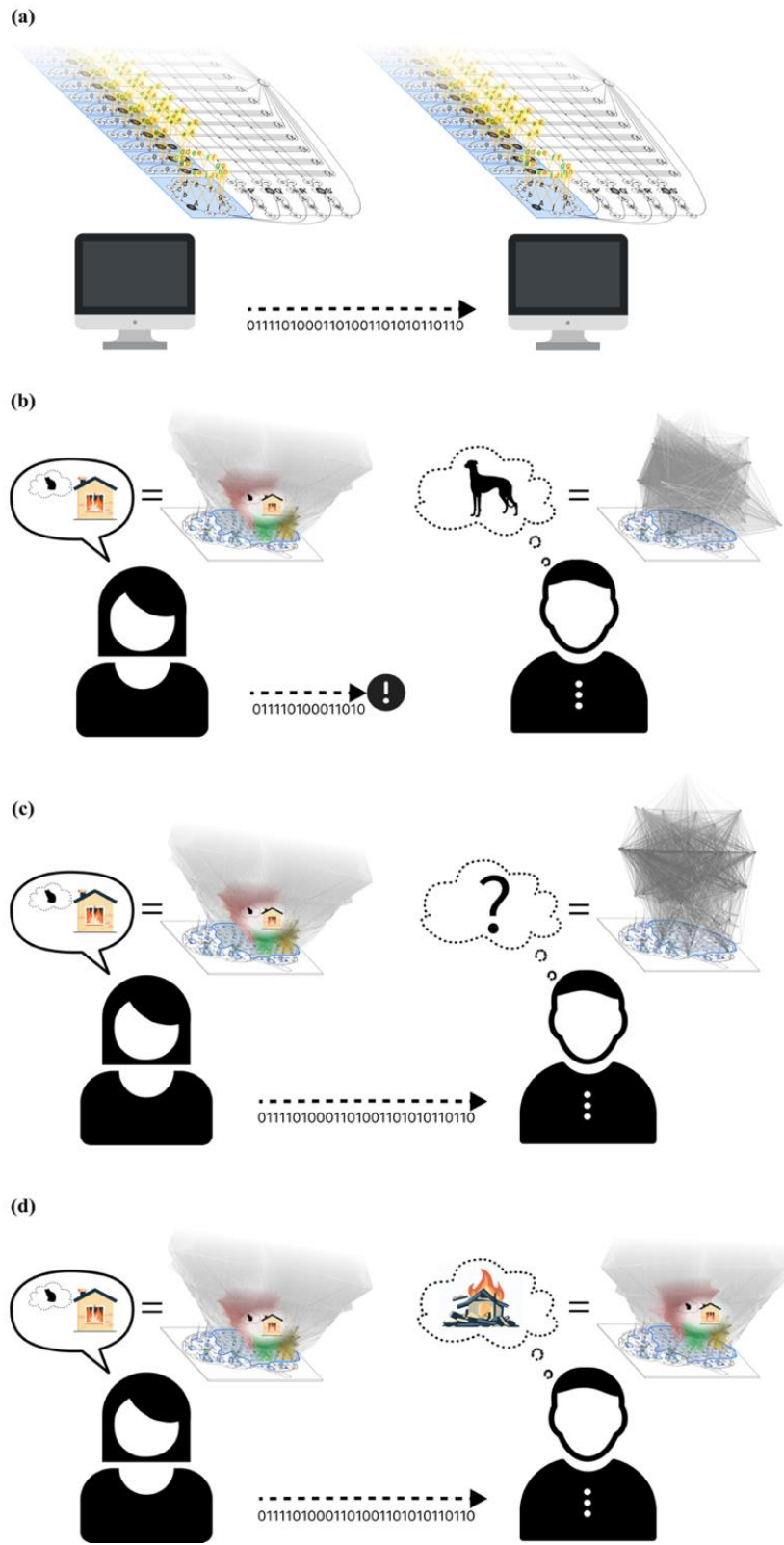

*Fig. 6. Communicating meaning according to IIT.* There is no communication of meaning if (a) the source or the target is not conscious, (b) the symbols are not transmitted successfully, or (c) the symbols do not trigger a similar sub-structure within the Φ-structure of the target. (d) Successful communication of meaning happens when all the above conditions are met, as reflected by the similarity of source and target Φ-folds.

Finally, Fig. 6(d) illustrates the successful communication of meaning. The source complex in its current state supports a $\Phi$-structure corresponding to the current experience, within which a $\Phi$-fold corresponds to a particular content. The complex's current state triggers an output over a motor interface, which sends a symbol over a communication channel. Through a sensory interface, the symbol triggers a state in the target complex that supports a large $\Phi$-structure, and a $\Phi$-fold within it bears structural similarity to the source's $\Phi$-fold. Appropriately, IIT's notion of information as the communication of meaning fits the original meaning of "informare" as "give form" to the mind.

Just as the $\Phi$ value of the $\Phi$-fold can be used to measure the amount of meaning, the similarity of the two $\Phi$-folds could be used to measure the amount of meaning communicated.[14] However, unlike Shannon information, which can be communicated perfectly in most instances, the communication of meaning will be approximate. This is because integrated information depends on the internal organization of source and target complexes, which can hardly be identical. Human brains share an evolutionary history, but developmental events and learning trajectories will necessarily result in individual differences in the precise wiring of the neural substrate of consciousness. Even if the activity pattern were the same, then, the intrinsic meaning of the $\Phi$-structure specified by each substrate would differ from person to person.[15]

## Conclusion

As outlined in this paper, both information theory[4] and integrated information theory (IIT)[5] take an axiomatic approach to characterize information and employ the tools of probability theory. However, they differ in critical ways: in essence, information theory deals with messages, IIT with meaning.

At the most general level, information theory is about the reliable transmission of symbols. It takes the extrinsic perspective of a communication engineer; characterizes information through axioms based on that perspective; picks a channel with its source, target, and capacity; and devises codes to optimally transmit messages across it.

IIT addresses the feeling/meaning of an experience. It takes the intrinsic perspective of a conscious subject and defines integrated information following axioms that characterize the essential properties of consciousness. In physical terms, this implies that the substrate of consciousness must be causal, intrinsic, specific, unitary, definite, and structured. Based on these properties, the IIT formalism can be used to unfold the intrinsic meaning of an activity pattern over a complex of units, yielding a $\Phi$-structure, and to quantify it as integrated information ($\Phi$). In principle, the communication of meaning can be assessed by measuring similarities between source and target $\Phi$-folds. Shannon information satisfies different properties, suited to quantifying the communication of symbols across a channel,

---

[14] This is not easy, as it requires unfolding structures with a very large number of components and developing ways to compare them optimally.

[15] A scenario not illustrated in the figure is a perfect one-to-one mapping in which every activity pattern (hence $\Phi$-structure) in the source complex would be able to send a different symbol to the target complex, where it would trigger a different $\Phi$-structure. However, the mapping would be between radically different $\Phi$-structures—that is, between radically different meanings (say, whenever the source thinks and says "gnat," the target hears and thinks "tang"). One could nevertheless find a transformation ("translation") that maximizes the communication of meaning. This case also emphasizes that, while the mapping may reveal symmetries among the meanings specified by the source and target complexes such that one "semantic space" could be rotated ("translated") into the other, in IIT's analysis, the meaning of an experience is defined in absolute terms by the distinctions and relations that compose a specific $\Phi$- structure, rather than by its similarity to other $\Phi$-structures. By the same token, similarities among $\Phi$-structures can be exploited to assess similarities among meanings in absolute terms.

but its formalism has nothing to say about the symbols' intrinsic meaning.

Unlike information theory, IIT was systematized only recently. While the core of the theory has remained unchanged, its formalism is still undergoing refinements and expansions[5, 45-47]. Furthermore, the exhaustive unfolding of $\Phi$-structures is unfeasible for large systems due to the combinatorial explosion of candidate systems, unit grains, distinctions, relations, and partitions. Nevertheless, the IIT framework has explanatory, predictive, and inferential power[12]. For example, it explains why certain regions of the brain can support consciousness while others cannot[12] and why consciousness is lost during dreamless sleep, anesthesia, and generalized seizures[17, 48]. Measures of complexity inspired by IIT can classify subjects as conscious vs. unconscious with unmatched sensitivity [18, 49]. IIT is also being used to account for the quality of experience—why space feels extended[15], time flowing[14], and objects as binding general concepts with particular features[13]—based on the organization of neural connections in different brain regions.

Finally, IIT's account of intrinsic meaning as a structure composed of integrated information has several implications. For example, while standard computers can be functionally equivalent to humans, their wiring is incompatible with consciousness[19]. Therefore, even though computers may engage in seemingly meaningful behaviors and utter seemingly meaningful sentences when driven by the proper software, their actions and utterances can only be imbued with intrinsic meaning when interpreted by a conscious being. Even among conscious entities, intrinsic meaning depends on the particular composition of $\Phi$-structures, which in turn depends on the precise wiring of their substrate, which is bound to vary as a result of evolutionary history, developmental events, and learning. Therefore, even shared meanings will differ to some extent, and many meanings will be idiosyncratic. This poses a limit on the communication of meanings but also suggests how one might proceed to assess it and improve it.

## Supplementary material I (information theory)

### Source entropy and source coding

If we know the probabilities of occurrence for all the symbols in the source alphabet, we can assign to each symbol a unique binary codeword. A message is then a sequence of binary codewords. Shannon showed that the expected length of the binary sequence generated by the source (the message) is minimized by assigning $log(1/p_n)$ binary digits to the $n^{th}$ symbol, where $p_n$ is its probability of occurrence[6]. This means that, given optimal *source coding*, the information content of a message is simply its length in bits. The minimum expected length per symbol (the rate at which information is produced) is called *source entropy*.

As a measure of information content, Shannon entropy $H = \Sigma_n\, p_n\, log\, (1/p_n)$ has the following desired properties (the first three can be used to derive it uniquely [4]):

Continuity: $H$ is continuous in $p_n$.

Additivity: If symbol generation is broken down to successive selections of symbols, the entropy of the overall process is the weighted sum of the entropies of the selection subprocesses. For example, if the source alphabet contains sounds in the English language, we can generate symbols by first choosing whether the symbol is a vowel or not, and then sample from the vowel/consonant subsets. The entropy of the overall sound-generation process is the entropy of the vowel/consonant selection plus the entropy of sampling from consonant and vowel subsets, weighted by their corresponding probabilities of selection.

Monotonic increase with number of outcomes: $H$ increases monotonically with the number of symbols $N$, if the symbols are equiprobable, $p_n = 1/N$.

Maximum: If the number of outcomes is fixed, $H$ is maximal when all the outcomes are equiprobable.

Symmetry: $H$ does not change if the outcomes are reordered. For example, if we generate the sequence using a new probability distribution, which uses a permutation of the original probabilities, the entropy value does not change.

These are natural properties if we define the information content of a message as the number of binary digits that minimizes its length. Importantly, it can be shown that Shannon entropy $H$ is a useful estimate of the number of bits required to represent the typical message from a source.[16]

### Mutual information, channel capacity, and channel coding

To assess how well information is transmitted across a channel, information theory resorts to *mutual information*, which characterizes the statistical dependence between probability distributions of outcomes (not individual outcomes) between source ($X$) and target ($Y$). Mutual information $I(X;Y)$, defined as $H(Y) - H(Y|X)$, can be rewritten as $D_{KL}(P(X,Y)\|P(X)P(Y))$—that is, the Kullback–Leibler divergence between the joint distribution and the product of the marginals. The measure is zero only when *X* and *Y* are independent. It is usually described as the information that random variable *Y* conveys about random variable *X* on average and vice versa.

---

[16] Thanks to the asymptotic equipartition property [6, 8, 50], if we have a distribution over $N$ symbols and draw $M$ samples from it, the probability of any typical occurrence is $2^{-MH}$, where $H$ is the Shannon entropy of the distribution and $M$ is large enough. This is because, with probability of almost 1, the $n^{th}$ symbol is going to occur $p_n M$ times in any outcome, and the probability of occurrence of any such outcome is then $\Pi_{n=1}^{N} p_n^{p_n M}$. If we want to represent these long equiprobable sequences with bits, we need to use at least $MH$ bits; otherwise we will run out of codewords to assign to all the $2^{MH}$ members. Thus, $MH$ is the minimum number of bits that we need to represent the sequences, and $H$ is the average number of bits per symbol.

Shannon defined the capacity of a channel as the maximum rate at which information can be transmitted through it, corresponding to the maximum of mutual information between source and target (the maximization is with respect to the source distribution). It can be shown that, for a long enough sequence of symbols, we can assign binary codewords to *X* such that, on average, $I(X;Y)$ bits can be recovered uniquely by observing *Y* [4].[17] Shannon further showed that, below channel capacity, it is possible to design an error control code (*channel coding*) whose probability of error is arbitrarily small.

## Information transmission and codes

In information theory, the most general definition of a code is a mapping from a set of symbols (such as outcomes of *X*) to another set (such as binary digits). Shannon's source coding allows one to recover (decode) the original symbols from their binary representations uniquely, while minimizing the expected length of the overall message (lossless compression). But in general, a code does not need to be binary, deterministic, or unique, or to minimize the expected length of messages. Thus, we can think of *Y* as a code for *X*, and of $P(Y|X)$ as the coding function, as long as $P(Y|X) \neq P(Y)$, i.e. $I(X;Y) > 0$ (that is, as long as we can answer *some* questions about *X* by observing *Y*). To illustrate, consider the simple channel in Fig. 7(a). Both *X* and *Y* ensembles contain four symbols, and each input symbol from *X* can lead to two output symbols in *Y* with equal probabilities. Both the input and output alphabets have four symbols, corresponding to 2 bits per symbol. However, the channel is noisy, its capacity is only 1 bit per symbol (regardless of the distribution over *X*, $H(Y|X)$, and the mutual information $I(X;Y)$ is at most 1 bit. Even so, *Y* can be considered as a code for *X* because observing *Y* allows us to answer at least some questions about *X*.

Moreover, which questions we can answer depends on extrinsic factors, namely the coding scheme determined by the channel designer. For example, the designer could choose to transmit only *r* and *g*, with equal probabilities, corresponding to a source entropy of 1 bit. The source could be encoded by assigning a 1-bit name to *r* and *g* (codebook {*r* : 0, *g* : 1}), achieving error-free communication at channel capacity. Or the designer could choose to achieve channel capacity by transmitting all the symbols with equal probability 0.25, corresponding to a source entropy of 2 bits. Then, using codebook {*r* : 00, *y* : 01, *g* : 11, *b* : 10} (Gray code), the receiver acquires full knowledge on 1 out of the 2 transmitted bits. Thus, when the receiver observes the yellow-green flag, they will know the second bit in the transmitted message is 1, while the first bit remains fully unknown. In short, mutual information tells us how many bits can be transmitted, but it does not tell us which bits, nor can it tell us what they mean—that is left entirely to the interpretation (codebook) of extrinsic observers.

---

[17] This is again based on the asymptotic equipartition property[6].

## Information processing and relevant codes

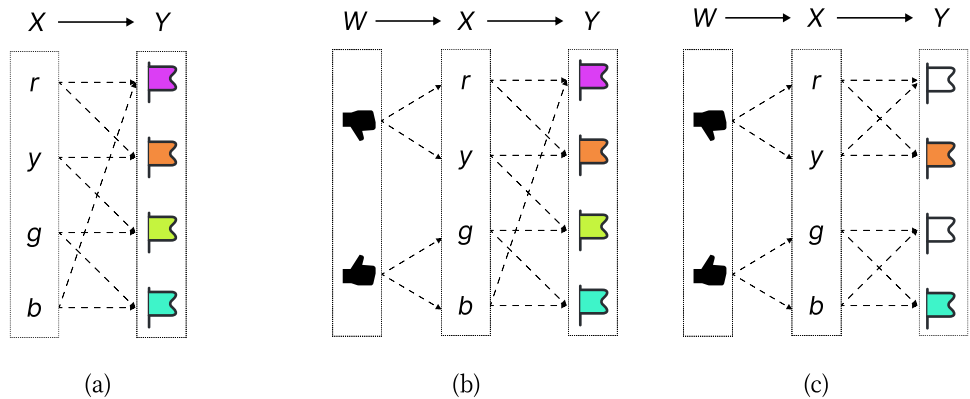

*Fig. 7. Codes and relevant codes. Examples showing how an external observer assigns meaning to codes through extrinsic partitioning of sample spaces. The arrows depict which symbols in the source lead to which symbols in the target with equal probabilities. (a) Due to the channel noise, the observer can at most recover 1 bit assigned to X by observing Y and can achieve this by designing a codebook for X and a decoding algorithm (see Supplementary material I). (b) The observer can introduce a new random variable W (say, representing the presence/absence of faces) that partitions the X space (space of visual stimuli) and define the relevant bit using W. While Y preserves 1 bit from X and X preserves 1 bit from W, Y only preserves 0.5 bit of the relevant information W. Y is still a code for W, a partitioning defined by the external observer. (c) The third channel preserves the relevant correlations, without changing the capacity, and makes Y an efficient code for W. Now, the observer is able to design a codebook for W and a decoding algorithm to recover the symbols assigned to W by observing Y.*

Extrinsic observers can not only choose particular codebooks but also extrinsic variables guiding the encoding and transmission of Shannon information. In other words, they can design *information-processing* channels to answer questions relevant to them, while ignoring other data[51].

For example, consider the information channels with two successive stages shown in Fig. 7(b) and Fig. 7(c). Suppose that we are not interested in recovering $X$ from $Y$, but only in a certain aspect of $X$, whose presence or absence we can encode with the random variable $W$. We can then define relevance as how well we can recover $W$ given $X$ vs. given $Y$. This is known as the information-bottleneck method and helps us to measure if $Y$ is efficiently squeezing information about $W$ from $X$[51]. This provides us with a framework to quantify not only how many questions we can answer about $X$ by observing $Y$, but which questions, as long as we can resort to an extrinsic variable $W$.[18]

In Fig. 7(b), $X$ is not the channel source but an intermediate encoding of $W$. The first stage of the channel, $W \to X$, does not introduce any noise. Channel capacity is 1 bit, so we can perfectly recover symbols of $W$, and the optimal input distribution is a uniform distribution over $W$. The second stage, $X \to Y$, also has a channel capacity of 1 bit. However, it does not preserve the relevant bit about $W$: $H(W) = 1$, $I(W;X) = 1$, but $I(W;Y) = 0.5$, meaning that we cannot recover $W$ exactly from $Y$ in one transmission. However, a trivial rearrangement of the channel can make $Y$ a more efficient code for $W$ without changing the channel capacity of $X \to Y$, as shown in Fig. 7(c). In this new scenario, $H(W) = 1$, $I(W;X) = 1$, and $I(W;Y) = 1$, leading to perfect recovery of $W$ (but not $X$) from $Y$. Thus, the question of whether $Y$ is a good code for $W$ boils down to which statistical dependencies are being preserved by $Y$. Information theory provides us with useful tools to examine such statistical dependencies. Even so, information processing can only decrease the information transmitted about the inputs to a channel and thus the number of questions we can answer about it[6, 8]. By the data-processing inequality, if we represent a channel as the Markov chain $W \to X \to Y$, then $H(W) \geq I(W;X) \geq I(W;Y)$. Most relevantly, information processing does not deal with how the meaning of the symbols would be assigned and interpreted.

---

[18] An interesting case arises when we define $W$ as the future state of $X$. In this case, what is relevant is how well the channel is able to predict the future (predictive processing)[25-27].

## Supplementary material II (IIT)

### Experience, its essential properties, and their operational formulation

IIT starts from experience itself, rather than from its behavioral, functional, or neural correlates[5, 10, 12]. The existence of experience, which is immediate and irrefutable, is the $0^{th}$ axiom of IIT.[19] From within consciousness, we can plausibly assume the existence of a world independent of our own experience. We can call this world *physical* if we can reliably observe it and manipulate it. IIT defines physical existence operationally as cause–effect power—the ability to "take and make a difference"—which is the $0^{th}$ postulate of IIT. Searching for the substrate of consciousness in physical terms then means formulating phenomenal existence (the $0^{th}$ axiom) operationally, in terms of cause–effect power (the $0^{th}$ postulate).

0) Existence: Experience *exists*. Therefore, the substrate of consciousness must be constituted of units that can *take and make a difference*.

IIT then identifies the essential properties of consciousness—immediate and true of every conceivable experience—as the five *axioms of phenomenal existence*. These properties are formulated in terms of cause–effect power as the five *postulates of physical existence*[5], which are employed to identify substrates of consciousness and unfold their structures. The five axioms and corresponding postulates are as follows:

1) Intrinsicality. Experience is *intrinsic*: it exists *for itself*. Thus, the cause–effect power of its substrate must be *intrinsic*: it must take and make a difference *within itself*.

2) Information. Experience is *specific*: it is *this one*. Thus, the cause–effect power of its substrate must be *specific*: it must be in *this state* and select *this cause–effect state*.

3) Integration. Experience is *unitary*: it is *a whole*, irreducible to separate experiences. Thus, the cause–effect power of its substrate must be *unitary*: it must specify its cause–effect state as *a whole set* of units, irreducible to separate subsets.

4) Exclusion. Experience is *definite*: it is *this whole—all of it*. Thus, the cause–effect power of its substrate must be *definite*: it must specify its cause–effect state as *this whole set* of units.

5) Composition. Experience is *structured*: it is composed of *distinctions* and the *relations* that bind them together, yielding a *phenomenal structure* that feels *the way it feels*. Thus, the cause–effect power of its substrate must be *structured*: subsets of units must specify cause–effects over subsets of units (*distinctions*) that can overlap with one another (*relations*), yielding a *cause–effect structure* (or Φ-*structure*) that is *the way it is*.

On this basis, IIT claims that every property of an experience must be accounted for by the *Φ*-structure specified by its substrate, with no additional ingredients. This is the fundamental *explanatory identity* of IIT[5].[20]

### Complexes

---

[19] In fact, experience is the starting point for everything, including science, logic, and mathematics[52].

[20] In addition, IIT employs ontological principles, key among them being the principles of maximal and minimal existence[5]. Maximal existence states that, when it comes to a requirement for existence, what exists is what exists the most. Minimal existence states that something cannot exist more than the least it exists.

Based on the postulates and the transition probability matrix (TPM) of a system of units, IIT proceeds to identify sets of units that qualify as substrates of consciousness (complexes)[5, 53]. A key role is played by the *intrinsic difference*[54] between probability distributions, a measure that uniquely satisfies the $0^{th}$ and first two postulates—being causal, intrinsic, and specific (much like Shannon's entropy function).

A complex is a set of units that has cause–effect power (existence) within itself (intrinsicality), is in a specific state and selects a specific cause and effect state over itself (information), and does so in a way that is not only irreducible (integration) but maximally so (exclusion). Irreducibility is measured by integrated information $\varphi_s$ ("system phi" or "small phi"), which for a complex must be higher than that of any set of units overlapping with it (by the principle of maximal existence). Any substrate of units condenses into a disjoint (non-overlapping) and exhaustive set of complexes, some of which may be exceedingly large and others negligibly small (Fig. 3). The "intrinsic units" of each complex also have a definite grain (typically, they will be macro-units constituted of many micro-units[55]), which is the one maximizing the existence of the complex. From the intrinsic perspective of a complex, the rest of the universe can be considered as a set of *background conditions* that can mediate extrinsic interactions.

### *Φ*-structures and differentiation capacity

Once a complex has been identified, IIT proceeds to unfold its *Φ*-structure (Fig. 3). This is composed of all the causal distinctions and relations specified by subsets of units within the complex. A subset of units specifies a *distinction* if it selects a specific, maximally irreducible cause and effect within the complex that is congruent (in the same state) with the cause and effect state of the complex as a whole. *Relations* obtain whenever there are overlaps among cause and/or effects of distinctions (details in [5]) Each distinction and relation has an associated value of integrated information ($\varphi$), whose sum is the structure integrated information (*Φ*, "structure Phi" or "big Phi"). The operation of determining all the distinctions and relations that compose the *Φ*-structure specified by a complex in a state is called *unfolding*.

The number of units in a complex and the organization of their intrinsic interactions impose an upper bound on the number and strength of the causal distinctions and relations the complex can specify in a given state[56]. This translates to an upper bound on the sum of the $\varphi$ values of all unique distinctions and relations a complex can specify across all its possible states, called differentiation capacity[11]. This bound is somewhat analogous to a channel's capacity as an upper bound on information transmission in the case of Shannon information.

### Contents of experience (*Φ*-folds)

For IIT, the *Φ*-structure accounts for both the quantity and quality of experience, with all its contents. The quantity is given by the *Φ*-structure's *Φ* value. The quality or way an experience feels—its "feeling"—is fully characterized by how the distinctions that compose the *Φ*-structure are related among themselves. The feeling of an experience is also its intrinsic meaning (*the feeling is the meaning*).

For example, consider the feeling of spatial extendedness, associated with most visual (and bodily) experiences, which is what "space" *means* from the intrinsic perspective of a subject. Phenomenal space can be characterized as a structure composed of "spots" that are related by reflexivity, inclusion, fusion, and connection[15]. Phenomenal structures of this kind can be accounted for by $\Phi$-structures specified by lattices of units, such as those found in posterior cortical areas of the brain, in line with neurological evidence[15]. More generally, any specific content within an experience, say, seeing the house on fire, would correspond to a part of the $\Phi$-structure—a $\Phi$-fold or sub-structure—composed of a highly interrelated subset of its distinctions and relations. This $\Phi$-fold corresponds to the feeling/meaning of seeing the house on fire from the intrinsic perspective of the subject of the experience (Fig. 5, right panel). For IIT, then, information is not a symbol or code but a structure, which corresponds to its intrinsic meaning. And information content is quantified not by the length of a message in bits but by the sum of the $\Phi$ values of the distinctions and relations composing the $\Phi$-fold.

**Meaning and matching**

Because the feeling/meaning of an experience is defined intrinsically, it does not matter how the experience is triggered. For example, the feeling/meaning of the house on fire is similar whether it is triggered by an external stimulus (Alice), imagined (Bob), or dreamt. However, one can assess the extent to which contents of experience are caused by external stimuli by multiplying the corresponding $\Phi$-folds by a triggering coefficient, yielding a measure of perceptual richness[11]. Accordingly, perceptions of external inputs can be considered as interpretations whose intrinsic meaning is provided by the $\Phi$-structure. In an adapted brain, intrinsic meanings will largely refer to (represent) relevant causal processes in the environment. For example, during wakefulness, correlated inputs due to exogenous causal processes triggers activity patterns that "up-select" (net strengthening) intrinsic connections, transforming extrinsic correlations into intrinsic causal powers. During sleep, intrinsic connections are "down-selected" (net weakening) based on activity pattern triggered endogenously. Over multiple sleep-wake cycles, the intrinsic connectivity is adjusted to preserve intrinsic meanings that "match" causal features of the environment at the expense of those that do not[11, 57, 58]. The overall matching between intrinsic meanings and causal processes in the environment can be estimated by measuring the differentiation of neural activity[11].